\documentclass[12pt]{iopart}

\usepackage{graphicx}
\usepackage{amssymb}
\usepackage{mathrsfs}
\usepackage[british]{babel}

\begin{document}

\title{How a finite potential barrier decreases the mean first passage time}

\author{Vladimir V. Palyulin$^{\dagger,1}$ and Ralf Metzler$^{\ddagger,\flat,2}$}
\address{$^{\dagger}$Physics Department, Technical University of Munich,
85747 Garching, Germany\\
$^{\ddagger}$Institute for Physics \& Astronomy, University of Potsdam,
14476 Potsdam-Golm, Germany\\
$^{\flat}$Physics Department, Tampere University of Technology,
FIN-33101 Tampere, Finland}
\ead{$^1$vladimir.palyulin@tum.de, $^2$rmetzler@uni-potsdam.de}

\date{\today}

\begin{abstract}
We consider the mean first passage time of a random walker moving in a
potential landscape on a finite interval, starting and end points
being at different potentials. From analytical calculations and Monte Carlo
simulations we demonstrate that the mean first passage time for a piecewise
linear curve between these two points is minimised by introduction of a
potential barrier. Due to thermal fluctuations this barrier may be crossed.
It turns out that the corresponding expense for this activation is less
severe than the gain from an increased slope towards the end point. In
particular, the resulting mean first passage time is shorter than for a
linear potential drop between the two points.
\end{abstract}


\maketitle

\section{Introduction}  

In classical mechanics, Bernoulli's 1696 \emph{brachistochrone\/} problem
addresses
the curve between two points that is covered by a point particle in the least
time, under the influence of gravity. If the particle starts at rest the
brachistochrone curve is a cycloid. Steeper at first, the particle is
accelerated, keeping its momentum in absence of friction. In particular at no
point along this curve the particle elevation is higher than that of the
starting point, for reasons of energy conservation. An overdamped, diffusing
particle may appear to behave classically: driven by a constant external force
the mean first passage time (MFPT) $T$ from one point to another along the
direction of the force equals $L/V$, the ratio of distance $L$ versus the
particle velocity $V$ \cite{Redner}. However, as the diffusing particle is
coupled to a heat bath, thermal fluctuations may lift it across a potential
barrier. At the same time, the overdamping does not allow the particle to
take along its momentum. To minimise the MFPT one would thus naively expect
that the particle should constantly move downhill. As we are going to show
here for the case of a piecewise linear potential, it is indeed beneficial
for the MFPT if the particle first crosses a potential barrier, that is,
the particle initially moves uphill. As a consequence the following downhill
slope becomes steeper, leading to a smaller overall MFPT.

\begin{figure}
\begin{center}
\includegraphics[width=7.2cm]{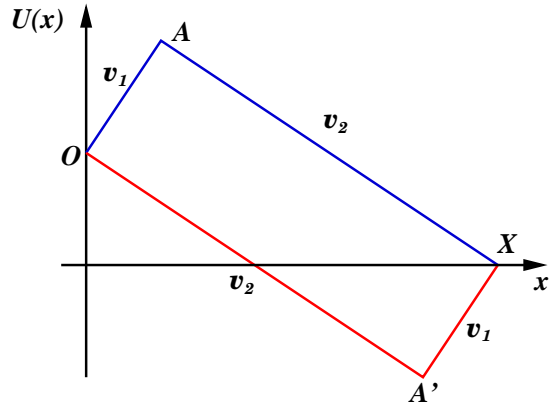}
\end{center}
\caption{Scheme of the piecewise linear potential (blue line) considered
here. The particle is initially placed at point $O$ (at $x=0$), in which we
impose a reflecting boundary condition. The end point is $X$, and we choose
$x_X=1$. At the turnover point $A$ the slope of the potential changes. $v_1$
and $v_2$ are the drift velocities on the two linear slopes ($v_1<0$ and $v_2
>0$). The red line shows the inversely symmetrical potential
resulting in the same MFPT (see text).}
\label{scheme}
\end{figure}

Generally, the question of the interplay between potential landscape and
diffusion properties is of great interest, resulting in often surprising
behaviour such as giant diffusivity \cite{GiantDiffusion1}. But which shape
of the potential should one choose in order to optimise the escape time on
an interval? A large number of previous studies were concerned with problems
of the escape from a potential well \cite{haenggirev}, following Kramers'
classical work \cite{Kramers}. Optimisation of the escape time may involve
phenomena such as resonant activation \cite{RA}. One of the simplest models
for a potential landscape
is a piecewise linear potential (Fig.~\ref{scheme}). Only recently it was
realised that an asymmetry in this kind of potential is important for escape
properties in resonant activation \cite{Wozinski,Spagnolo}. The asymmetry of
the potential also plays a crucial role in systems with periodic potentials
relevant to molecular motor models \cite{Motor,Motor1,Motor2}, or for
molecular shuttles in suprachemical compounds \cite{shuttle}. However, to the
best of our knowledge the role of asymmetry on the MFPT for a static potential
as displayed in Fig.~\ref{scheme} has not been discussed.

\section{Mean first passage time minimisation}

We consider a particle diffusing from the starting point $O$ at $x=0$, to point
$X$ located at $x_X=1$, in the piecewise linear potential going through point
$A$ at $x_A$. This situation is sketched in Fig.~\ref{scheme}. The values of the
potential in these points are $U_O$, $U_A$, and $U_X=0$, without loss of
generality. At the starting point $O$ we impose a reflecting boundary condition
while at the end point $X$ we apply an absorbing boundary condition for the
calculation of the MFPT. The question we pursue is: which shape of the
piecewise linear potential minimises the MFPT from $O$ to $X$?

The MFPT for the piecewise linear potential with bias velocities $v_1$ (on $0
\le x\le x_A$) and $v_2$ (on $x_A<x<1$) on the unit interval, shown in
Fig.~\ref{scheme}, readily obtains analytically \cite{Redner,Privman1}.
A unit current $j(0,t)=\delta(t)$ is injected at $x=0$, and the output is
calculated from the solution of the Fokker-Plank equation,
\begin{equation}
\label{fpe}
\frac{\partial P(x,t)}{\partial t}=\left(\frac{\partial}{\partial x}\frac{U'(x)}
{m\eta}+D\frac{\partial^2}{\partial x^2}\right)P(x,t),
\end{equation}
where $U'(x)$ is the derivative of the external potential. Moreover $m$ is the
particle mass, $\eta$ the friction experienced by the particle, and $D$ is its
diffusion constant. For the gravitational potential $U(x)=mgh(x)$ for a particle
at elevation $h(x)$ at position $x$ and with the gravitational constant $g$,
the drift term in the Fokker-Planck equation becomes $\partial/\partial x\left(
gh'(x)/\eta\right)P(x,t)$. The ratio $g/\eta$ has the dimension of a velocity,
so that the Fokker-Planck equation may be rewritten in the form
\begin{equation}
\frac{\partial P(x,t)}{\partial t}=\left(-v_i\frac{\partial}{\partial x}+
D\frac{\partial^2}{\partial x^2}\right)P(x,t),
\end{equation}
with piecewise constant drift velocity $v_i$, where $i=1,2$ denotes the two
domains with piecewise linear potential. Note the sign of the drift velocity:
an increase of the potential causes a drift to the left, and \emph{
vice versa}. The reflecting and absorbing boundary conditions at $x=0$ and
$x=1$, respectively, read $\left.\frac{\partial P}{\partial x}\right|_{x=0}$
and $P(1,t)=0$. Requiring continuity of the distribution $P$ and the probability
flux at point $A$, the MFPT yields in the form \cite{Redner}
\begin{eqnarray}
\nonumber
T&=&\frac{D}{v_{1}v_{2}}\left(1-e^{-v_{1}x_A/D}\right)\left(1-e^{-v_{2}
(1-x_A)/D}
\right)\\
\nonumber
&&+\frac{x_A}{v_{1}}+\frac{1-x_A}{v_{2}}+\frac{D\left( e^{-v_1 x_A/D}-1\right)}
{v_1^2}\\
&&+\frac{D\left(e^{-v_2(1-x_A)/D}-1\right)}{v_2^2},
\label{mfpt}
\end{eqnarray}
as function of $x_A$, $v_1$, and $v_2$.
We note that all variables occurring in Eqs.~(\ref{fpe}) to (\ref{mfpt}) are
dimensional. In what
follows we measure lengths in units of cm and time in sec. Thus when writing
$L=1$ for the distance between starting and end points, this actually means
1cm.

Let us study the MFPT (\ref{mfpt}) in detail. We first note that expression
(\ref{mfpt}) is symmetric under simultaneous exchange of $v_1\leftrightarrow
v_2$ and $x_A\leftrightarrow 1-x_A$, i.e., inversion through the midpoint of
the line connecting $O$ and $X$. This inverse case corresponds to the red line
in Fig.~\ref{scheme}. Secondly, we observe that by increasing the elevation of
point $A$ with respect to $O$ and $X$ and shifting the turnover point $A$
towards the starting point $O$ such that $|v_1x_A|\gg1$, $|v_2|(1-x_A)\gg1$,
and $x_A\ll1$, the MFPT (\ref{mfpt}) reduces to
\begin{equation}
\label{LargeAmplitude}
T\approx\frac{D}{v_1^2}e^{|v_1x_A|/D}+\frac{1}{v_{2}}.
\end{equation} 
This is the sum of the MFPTs on the two subintervals. Indeed, the first term
corresponds to the Kramers rate for crossing of a high potential barrier,
see below, while the second term represents the MFPT at constant drift
$v_2$ over the unit distance. Result (\ref{LargeAmplitude}) demonstrates that
the the overall
MFPT $T$ as well as both individual terms are reduced by increase of $A$'s
elevation while keeping the product $v_1x_A$ constant. This is one of the
central results of our study: the introduction of a high but narrow barrier 
reduces the MFPT.

For the thermally activated crossing of a sufficiently high potential barrier
the corresponding barrier crossing time was obtained by Kramers
\cite{Kramers,Risken},
\begin{eqnarray}
\label{kramers}
T_K=\frac{2\pi}{\sqrt{U''(x_{\mathrm{min}})|U''(x_{\mathrm{max}})|}}
e^{\left[U(x_{\mathrm{min}})-U(x_{\mathrm{max}})\right]/D}.
\end{eqnarray}
Here $x_{\mathrm{min}}$ and $x_{\mathrm{max}}$ denote the positions of the
potential minimum (where the particle is initially placed) and the saddle
of the potential. According to expression (\ref{kramers}) this characteristic
time depends both on the potential difference $\Delta U=U(x_{\mathrm{max}})
-U(x_{\mathrm{min}})$ and the curvature of the potential at these two points.
If we imagine that we smoothen the piecewise linear potential around the
minimum and maximum points, it becomes clear that for fixed $\Delta U$ a
decrease of the distance between $x_{min}$ and $x_{max}$ implies an increase
of the respective curvatures and thus a \emph{decrease\/} of the barrier
crossing time. This observation underlines that our above result for the MFPT
in the piecewise linear potential is consistent with the physics of barrier
crossing.

What happens in the case opposite to Eq.~(\ref{LargeAmplitude}), when the two
drift velocities are small, $|v_1|,|v_2|\ll1$? Expansion of Eq.~(\ref{mfpt})
up to first order yields
\begin{eqnarray}
\nonumber
T&\approx&\frac{1}{2D}-\frac{v_1 x_A^2}{6D^2}\left(1+2(1-x_A)\right)\\
&&-\frac{v_2(1-x_A)^2}{6D^2}(1+2x_A).
\label{low_v_limit}
\end{eqnarray}
Here, the first term represents the MFPT of free diffusion on the unit interval.
The next two terms are the first order corrections in $v_1$ and $v_2$. Depending
on the actual values of $v_1$ and $v_2$ these terms may either lead to a
decrease or increase of the MFPT.

\begin{figure}
\begin{center}
\includegraphics[width=10 cm]{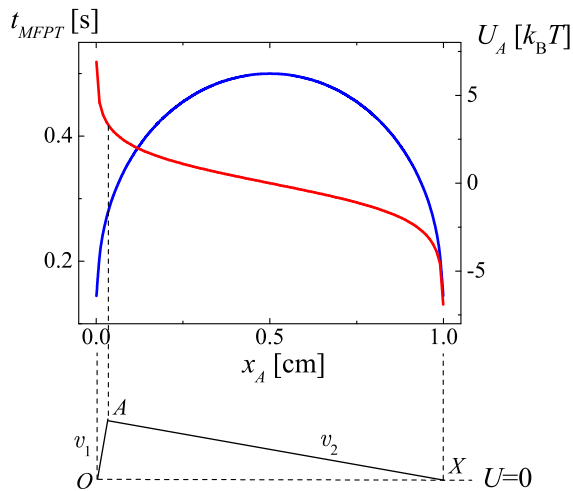}
\end{center}
\caption{Minimal MFPT in the piecewise linear potential for vanishing potential
difference between starting and end point, as function of the turnover point
position $x_A$ (blue curve). The corresponding optimum value for the value of
the potential at the turnover point is shown as the red line. The dashed line
emanating from the turnover point in the schematic of the potential profile
(bottom of graph) intersects the two curves at the associated values of MFPT
and $U_A$.}
\label{deltaUzero}
\end{figure}

\begin{figure}
\begin{center}
\includegraphics[width = 10 cm]{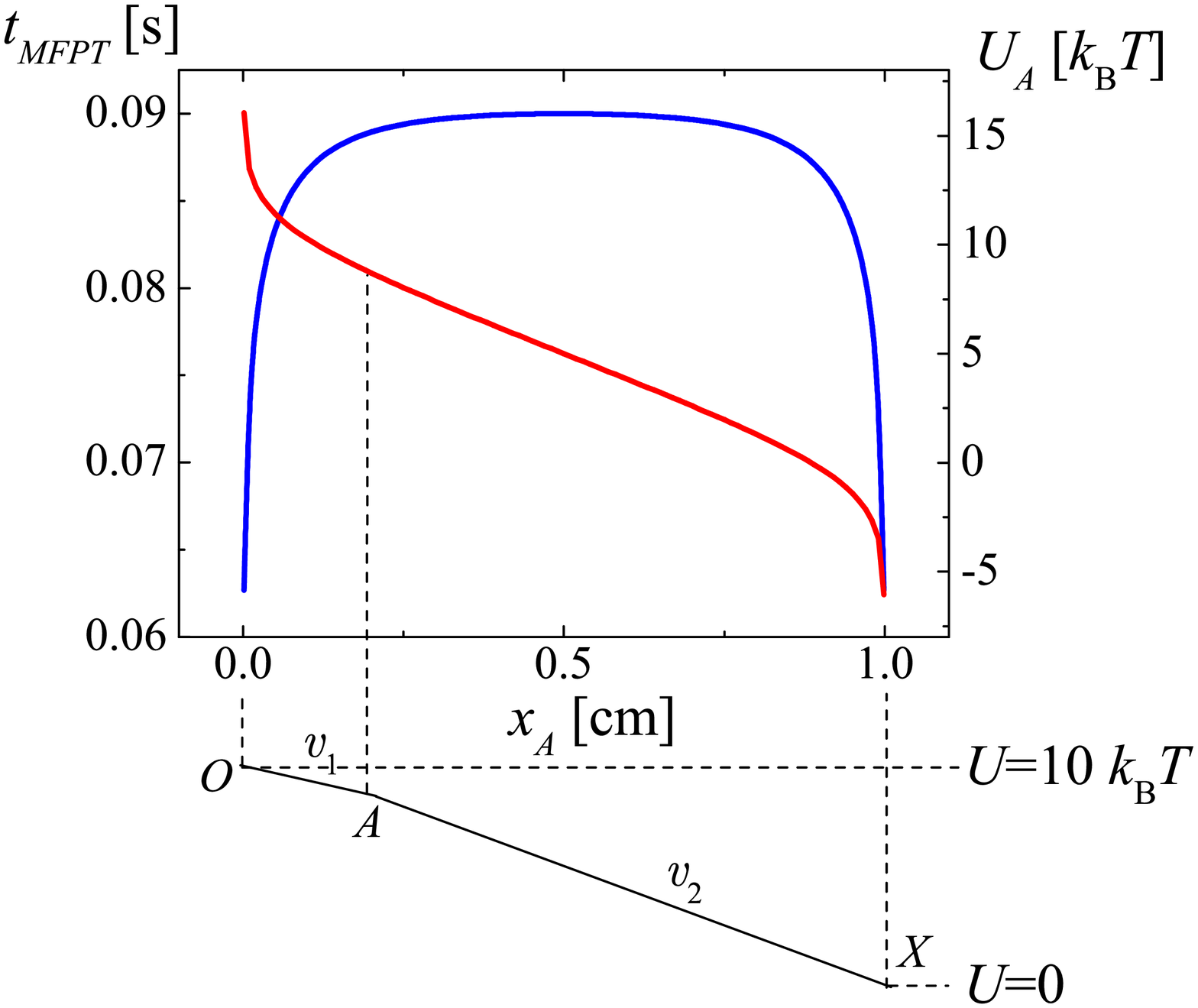}
\end{center}
\caption{Minimal MFPT and corresponding height of the potential at the
turnover point $A$ as function of the position $x_A$, in the case when
the potential difference between starting and end points is 10 $k_BT$.}
\label{nonzeroU_Acurve}
\end{figure}

\begin{figure}
\begin{center}
\includegraphics[width = 10 cm]{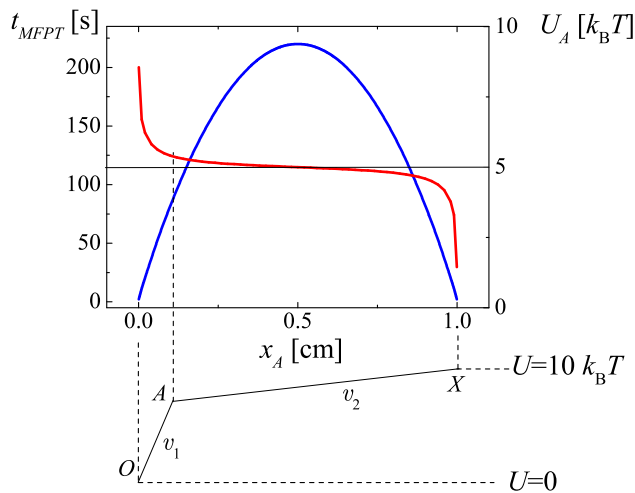}
\end{center}
\caption{Minimal MFPT and associated turnover potential for the case when
the potential difference between starting and finishing points is $-10k_BT$
(end point is higher than the starting point).}
\label{BarrierClimb}
\end{figure}

While the MFPT can be arbitrarily reduced by increasing $v_1$ (and thus also
$v_2$) and simultaneously decreasing the position $x_A$ of the turnover point,
a finite potential barrier may still reduce the MFPT. We analyse the three
possible,
different cases in Figs.~\ref{deltaUzero}-\ref{BarrierClimb}. Starting with
the case when starting and end points are at the same potential level, in
Fig.~\ref{deltaUzero} we show the minimal value for the MFPT (\ref{mfpt})
together with the corresponding optimal value for the potential at $A$, $U_A$,
as function of the position $x_A$ of the turnover point. This minimisation was
performed numerically with Mathematica. We see that the largest value of the
MFPT is obtained when the turnover point is located in the middle of the
interval at $x_A=0.5$. In this special case the optimum is reached in
absence of a potential barrier ($U_A=0$), i.e., for unbiased
diffusion. Away from the midpoint, the MFPT appears dramatically reduced.
For $x_A\to0$ and $x_A\to1$, the fastest MFPT is obtained when the potential
diverges, $U_A\to\pm\infty$. Notice the symmetries of both MFPT and profile
of optimal turnover points with respect to the midpoint, $x_A=0.5$.

For the case of very asymmetric positions of turnover points $x_A\to0$, the
optimal value for the drift $v_1$ can be computed analytically, if the
potential difference $\Delta U=v_1x+v_2(1-x)$ and $x_A$ are fixed. Expansion
of expression (\ref{mfpt}) as a series for small $x_A$ leads to the first
order approximation
\begin{eqnarray}
T\approx\frac{1}{\Delta U}+\frac{D\Omega_-}{\Delta U^2}
-\frac{D(\Delta U-v_1)\left(\frac{\Delta U}{D}\Omega_++2\Omega_-\right)}{\Delta
U^3}x_{A},
\label{smallxAexp}
\end{eqnarray}
where $\Omega_{\pm}=\exp(-\Delta U/D)\pm1$. Here the first two terms are the
MFPT for a uniform linear bias with potential difference $\Delta U$. The third
term is the correction linear in $x_A$. Analysing its form shows that an
increase of the height of the turnover point (i.e., an increase of $|v_1|$)
always
leads to a decrease of the MFPT if $\Delta U$ is positive. For the optimal
slope $v_1$ we obtain the approximate expression 
\begin{eqnarray}
v_{1}\approx-\frac{\Delta U}{2x_{A}}\frac{\left(\frac{\Delta U}{D}\Omega_+
+2\Omega_-\right)(2-4x_A-\frac{\Delta U}{D}x_{A})}{\left(6\Omega_-+\frac{\Delta
U^2}{D^2}e^{-\frac{\Delta U}{D}}+2\frac{\Delta U}{D}
(1+2e^{-\frac{\Delta U}{D}})\right)}.
\label{analyticV1}
\end{eqnarray}
In the range of small $x_A$ and $\Delta U>0$ all terms in the brackets are
positive. Hence, expression (\ref{analyticV1}) proves analytically that
in this case a barrier indeed optimises the MFPT. Note that the numerical
accuracy of this approximation is actually not too good. In order to reproduce
the functional behaviour over a longer range of $x_A$ higher order terms need
to be considered.

For the case when the starting point is higher than the end point, the result
for the minimal MFPT is displayed in Fig.~\ref{nonzeroU_Acurve}. Here the MFPT
shows an extended plateau around $x_A=0.5$. Exactly at this midpoint the
minimum MFPT corresponds to the naively expected case of a constant slope from
starting to end point. For $x_A$ closer to zero the MFPT again drops down to
zero while the value of the potential at the turnover point diverges. Both
curves for the MFPT and the potential at the turnover point are again symmetric
with respect to the midpoint. In contrast to Fig.~\ref{deltaUzero}, however,
the curve for the MFPT is not symmetric around the zero-line of the potential.

For completeness we consider the case when the end point is elevated with
respect to the starting point. While a classical particle would never reach
this end point, a thermally driven particle may gain the necessary energy
from the heat bath. The corresponding optimal potential of the turnover
point in the piecewise linear potential and the associated MFPT are shown
in Fig.~\ref{BarrierClimb}. It turns out to be beneficial when an initial
barrier exists whose height exceeds the overall potential difference
$|\Delta U|$ between starting and end point, such that the drift velocity
$v_2$ is positive.

Let us compare the minimal MFPT in the three cases of positive, zero, and
negative potential difference between the initial and end points of our
setup, for $x_{A1}=10^{-3}$ and $x_{A2}=0.5$ (i.e., the longest MFPT). For
$\Delta U=10k_BT$ (Fig.~\ref{nonzeroU_Acurve}) the ratio $T(x_{A1}):T(x_{A2})
\approx0.7$, for $\Delta U=0$ (Fig.~\ref{deltaUzero}) it is $T(x_{A1}):T(x_{A2})
\approx0.29$, and for $\Delta U=-10k_BT$ (Fig. \ref{BarrierClimb}) we find
$T(x_{A1}):T(x_{A2})\approx0.01$. Thus, the introduction of a potential barrier
or kink has indeed the largest effect on the MFPT when the end point has a
\emph{higher\/} energy. That is, when it is harder to reach the end point
energetically, the benefit from a potential turnover is larger. This is the
second central result of our study.

We simulated the Brownian motion of a particle in a piecewise linear potential
with a Monte Carlo approach, based on the Metropolis algorithm: If the potential
difference $\delta U$ between current and potential new position is positive,
$\delta U>0$, then the step is accepted with probability $\exp\left(-\delta U/
[k_BT_M]\right)$, where $k_BT_M$ is a measure of temperature. Otherwise the step
is immediately accepted.

Comparison with the analytical results was achieved by consideration of the
continuum limit of a discrete biased random walk on a lattice. The probability
distribution of jumps of length $\ell$, $p(\ell)$, is defined by the
Fokker-Planck equation \cite{Hughes}
\begin{equation}
\label{discreteFP}
\frac{\partial c(x,t)}{\partial t}=-\frac{\Delta}{\tau}m_1\frac{\partial
c(x,t)}{\partial x}+\frac{\Delta^2}{2\tau} m_2 \frac{\partial^2 c(x,t)}{
\partial x^2}
\end{equation}
where it is assumed that the lattice spacing and time step are infinitely
small: $\Delta\rightarrow0$, $\tau\rightarrow0$, and $m_1=\sum\ell p(\ell)$,
$m_2=\sum\ell^2 p(\ell)$. Hence,
\begin{equation}
D= \lim_{\Delta,\tau\rightarrow 0}\frac{m_2\Delta^2}{2\tau},\,\,\,
v= \lim_{\Delta,\tau\rightarrow 0}\frac{m_1\Delta}{\tau}
\end{equation}
In the case we considered, the values of diffusion constants and the slopes
in continuum limit are
\begin{equation}
D\approx\frac{1}{2N^2\tau},\,\,
v_1\approx\frac{U_A}{2x_AN^2k_BT_M\tau},
\,\,v_2\approx\frac{U_A}{2(1-x_A)N^2k_BT_M\tau},
\end{equation}
where $N$ is the lattice size and $x_A$ the position of the turnover point.

\begin{figure}
\begin{center}
\includegraphics[width = 9 cm]{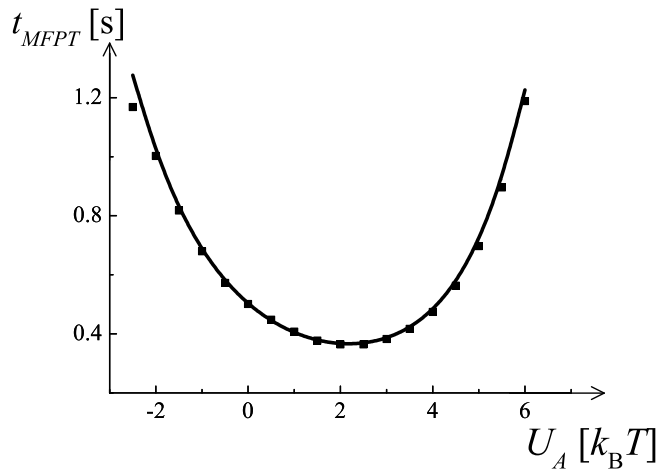}
\end{center}
\caption{Comparison of Monte Carlo simulations (squares) with the analytical
result from Eq.~(\ref{mfpt}) shown as the full line. The lattice size is $N=
1,001$, and the number of runs is 100,000.}
\label{MC}
\end{figure}

The simulations demonstrate excellent agreement with our analytical results.
We show the comparison between simulations and Eq.~(\ref{mfpt}) for the case
$\Delta U=0$ for $x_A=0.1$, $k_BT_M=1$, and $N=1001$ in Fig.~\ref{MC}.

\section{Discussion}

On a flat potential landscape significant progress has been achieved in the
theory of MFPTs on arbitrary, finite domains \cite{olivier1}. In particular,
the role of compact versus non-compact explorations has been revealed in
generality \cite{olivier2}. Much less is known about MFPT properties in
potential landscapes.

We analysed the value of the MFPT in a finite interval for a piecewise linear
potential, finding that the introduction of a barrier reduces the MFPT. In the
ideal case when the barrier height is unlimited the MFPT can be reduced
arbitrarily. This \emph{a priori\/} surprising results were shown to be in
line with physical arguments such as Kramers escape theory, and may be of
interest in the design of potential energy landscapes, for instance, for
functional molecules (molecular shuttles), or for molecular motors. Conversely,
our results may shed new light on the role of barriers in known landscapes,
for instance, in the folding landscape of proteins. Indeed, it was shown in
Ref.~\cite{wagner} that intermediate barriers of height $>1k_BT$ increase the
folding rate of proteins \cite{wagner}.

\begin{figure}
\begin{center}
\includegraphics[width= 8 cm]{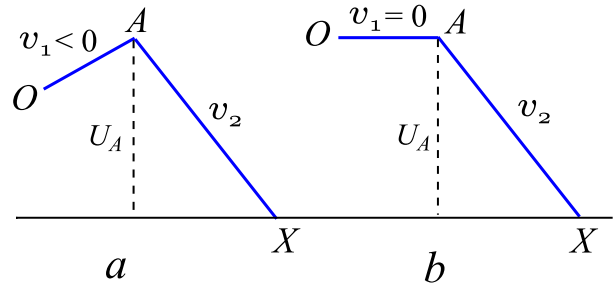}
\end{center}
\caption{The MFPT in case (a) is larger than in case (b) as long as $U_A$
is fixed.}
\label{U_limited}
\end{figure}

\begin{figure}
\begin{center}
\includegraphics[width=8cm]{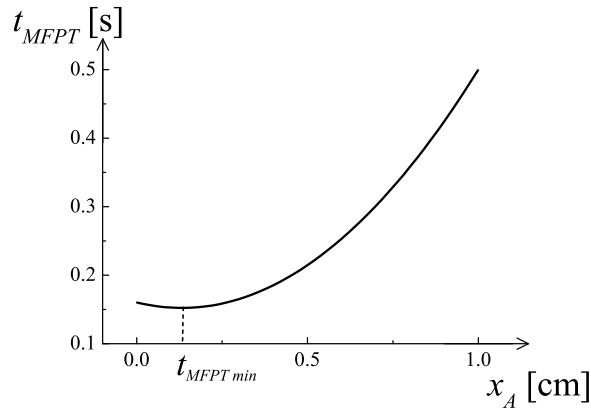}
\end{center}
\caption{MFPT for the case of fixed potential $U_A$ corresponding to
Fig.~\ref{U_limited}(b) as function of the position $x_A$ of the turnover
point. Here, $\Delta U=5 k_BT$.}
\label{xVariation}
\end{figure}

What happens if the height of the potential barrier is limited? Consider
the situation sketched in Fig.~\ref{U_limited}. If the values of $U$
at the starting and end points of the interval are fixed and the height of
the potential in point $A$ is fixed, it is clear that in case (a) the MFPT
is higher than in case (b). This changes considerably the answer to the MFPT
minimisation task. Starting with a horizontal slope we could still imagine that
a shift of the turnover point $A$ may optimise the MFPT: if shifted to the
right we have an increase for the time to reach $A$ but a gain from an
increased drift velocity $v_2$. Variation of $x_A$ in this case leads to
the dependence shown in Fig.~\ref{xVariation}. At the right end of the interval
between starting and end points the behaviour tends to the value $T=0.5$,
corresponding to unbiased diffusion.
The gain at the optimum value for $x_A$
in this case is in fact only a few per cent, compared to the case of a linear
potential drop ($x_A=0$).

All results presented above demonstrate the critical importance of the asymmetry
of the potential barrier for optimisation of the MFPT. This gain rests on the
significant facilitation of the passage on the long easy slope which
overcompensates losses for crossing of the barrier. The result (\ref{mfpt})
allows the adjustment of the MFPT to any finite value, including infinitely
large and infinitely small times. However, if one wants to decrease the MFPT to
some specific, small value, this result also shows that, to compensate an
increase in barrier height, a substantial reduction of the position $x_A$ of
the turnover point is required.

In classical mechanics the cycloid is the optimal curve for a point particle
in absence of friction: after an initial steep descent, i.e., high acceleration,
the momentum of the particle carries on. For a diffusing, overdamped particle
in the case of piecewise linear potential provide the answer is qualitatively
the opposite: in order to minimise the MFPT there should be a steep and short
ascent.

It will be interesting to consider more complex shapes of the potential, in
particular, the case of multiple barriers as mentioned in the context of
protein folding \cite{wagner}. Moreover, numerical analysis of the first
passage distribution associated with the process considered herein will be
of interest, as well as the consideration of the full motion including
inertia effects.
Although it is possible to optimise the potential by trial and error for a
fixed set of potential shapes, the question about whether the optimisation
algorithm exists in generality, remains to be investigated.
Another interesting question
is whether similar results could be obtained under anomalous diffusion
conditions \cite{report}.

\ack
VVP wishes to acknowledge financial support from Deutsche Forschungsgemeinschaft,
and Vladimir Yu. Rudyak for discussions about algorithms and randomisation. RM
acknowledges support from the Academy of Finland within the FiDiPro scheme. The
authors would like to thank the anonymous referee for pointing out the
interesting Ref.~\cite{wagner}.


\section*{References}

\begin{thebibliography}{99}

\bibitem{Redner} S. Redner, A Guide to First-Passage Processes (Cambridge
University Press, Cambridge, UK, 2001).

\bibitem{GiantDiffusion1} P. Reimann, C. Van den Broeck, H. Linke,
P. H{\"a}nggi, J. M. Rubi, and A. Pérez-Madrid, Phys. Rev. E \textbf{65},
031104 (2001); Phys. Rev. Lett. \textbf{87}, 010602 (2001).

\bibitem{haenggirev} P. H{\"a}nggi, P. Talkner, and M. Borkovec,
Rev. Mod. Phys. \textbf{62}, 251 (1990).

\bibitem{Kramers} H. A. Kramers, Physica \textbf{7}, 284 (1940).

\bibitem{RA} C. R. Doering and J. C. Gadoua, Phys. Rev. Lett. \textbf{16},
2318 (1992); L. Gammaitoni, P. H{\"a}nggi, P. Jung, and F. Marchesoni,
Rev. Mod. Phys. \textbf{70}, 223 (1998).

\bibitem{Wozinski} A. Wozinski and J. Iwaniszewski, Phys. Rev. E \textbf{80},
011129 (2009).

\bibitem{Spagnolo} A. Fiasconaro and B. Spangnolo, Phys. Rev. E \textbf{83},
041122 (2011).

\bibitem{Motor} P. Reimann, Phys. Rep. \textbf{361}, 57 (2002).

\bibitem{Motor1} M. Porto, M. Urbakh, and J. Klafter, Phys. Rev. Lett.
\textbf{85}, 491 (2000).

\bibitem{Motor2} G. Oshanin, J. Klafter, and M. Urbakh, Europhys. Lett.
\textbf{68}, 26 (2004).

\bibitem{shuttle} R. A. Bissell, E. C{\'o}rdova, A. E .Kaifer, and J. Fraser
Stoddart, Nature \textbf{369}, 133 (1994).

\bibitem{Privman1} H. L. Frisch, V. Privman, C. Nicolis, and G. Nicolis,
J. Phys. A \textbf{23}, L1147 (1990); V. Privman and H. L. Frisch, J. Chem.
Phys. \textbf{94}, 8216 (1991).

\bibitem{Risken} H. Risken, The Fokker-Planck Equation (Springer Verlag,
Berlin, 1989).

\bibitem{Hughes} B. R. Hughes, Random Walks and Random Enviroments, Vol.1:
Random Walks (Clarendon Press, Oxford, UK 1995).

\bibitem{olivier1} S. Condamin, O. B{\'e}nichou, V. Tejedor, R. Voituriez,
and J. Klafter, Nature \textbf{450}, 77 (2007).

\bibitem{olivier2} O. B{\'e}nichou, C. Chevalier, J. Klafter, B. Meyer, and
R. Voituriez, Nature Chem. \textbf{2}, 472 (2010).

\bibitem{wagner} C. Wagner and T. Kiefhaber, Proc. Natl. Acad. Sci. USA
\textbf{96}, 6716 (1999).

\bibitem{report} R. Metzler and J. Klafter, Phys. Rep. \textbf{339}, 1
(2000).

\end{thebibliography}
\end{document}